\documentstyle[12pt]{article}
\begin{document}

\def\zbar{{\bar z}}
\def\half{{1 \over 2}}
\def\parz{\partial_z}
\def\parzbar{\partial_{\bar z}}
\def\bfv{{\bf v}}
\def\bfx{{\bf x}}
\def\co{{\cal O}}
\def\e{{\rm e}}

\begin{flushright}
OCHA-PP-48 \ \ \ \ \\
NDA-FP-16\ \ \ \ \ \ \ \\
September 1994
\end{flushright}

\vfill

\begin{center}
{\large\bf
Quantum Mechanics for the Swimming \\
of Micro-Organism in Two Dimensions\\
}

\vfill

{\normalsize
Shin'ichi NOJIRI$\ ^{*)}$, Masako KAWAMURA and Akio SUGAMOTO}

\vfill

{\it
Department of Physics, Faculty of Science,
Ochanomizu University

1-1, Otsuka 2, Bunkyou-ku, Tokyo, 112, JAPAN

\vfill

$*)$ Department of Mathematics and Physics

National Defence Academy, Yokosuka, 239, JAPAN}

\end{center}

\vfill

\begin{abstract}
In two dimensional fluid, there are only two classes of swimming ways
of micro-organisms, {\it i.e.}, ciliated and flagellated motions.
Towards understanding of this fact, we analyze the swimming problem
by using $w_{1+\infty}$ and/or $W_{1+\infty}$ algebras. In the study
of the relationship between these two algebras, there appear the wave
functions expressing the shape of micro-organisms. In order to
construct the well-defined quantum mechanics based on $W_{1+\infty}$
algebra and the wave functions, essentially only two different kinds
of the definitions are allowed on the hermitian conjugate and the
inner products of the wave functions. These two definitions are
related with the shapes of ciliates and flagellates. The formulation
proposed in this paper using $W_{1+\infty}$ algebra and the wave
functions is the quantum mechanics of the fluid dynamics where the
stream function plays the role of the Hamiltonian. We also consider
the area-preserving algebras which arise in the swimming problem of
micro-organisms in the two dimensional fluid. These algebras are
larger than the usual $w_{1+\infty}$ and $W_{1+\infty}$ algebras. We
give a free field representation of this extended $W_{1+\infty}$
algebra.
\end{abstract}

\newpage

\section{Introduction}

It is known that there exist only three different universality
classes of the swimming ways of micro-organisms;  (1) Swimming with
cilia is adopted by the spherical organisms with the length scale of
$20 \sim 2 \times 10^4 \mu$m, an example of which is
{\it paramecium}; (2) the smaller micro-organisms with the size of
$1 \sim 50 \mu$m swim with flagella, an example of which is the
{\it sperm}; (3) the bacteria with the size of $0.2 \sim 5 \mu$m swim
with bacterial flagella the motion of which resembles the screwing of
the wine-opener. The swimming of micro-organisms was analysed from
the gauge theoretical viewpoint by Shapere and Wilczek \cite{3}. The
problem was also studied in our previous papers \cite{1} from the
viewpoint of string and membrane theories. Our target is still the
understanding why it is possible that such simple classification is
realized in the swimming problem of micro-organisms. It might be
possible to expect that the answer would be obtained by clarifying
the algebraic structure of the swimming motion.

Recently there were much progresses in the representation theory of
$W_{1+\infty}$ algebra \cite{4,5,6,7,8,9}. By using the
$w_{1+\infty}$  and/or $W_{1+\infty}$ algebras, we analyze in this
paper the swimming problem of the micro-organisms in two dimensions.
In two dimensions, there are two classes of the swimming ways,
{\it i.e.}, ciliated and flagellated motions. In a representation of
the algebra, there appear the wave functions expressing the shape of
micro-organisms. In order to give a consistent quantum mechanics
based on $W_{1+\infty}$ algebra and the wave functions, essentially
only two different kinds of definitions of the hermitian conjugate
and the inner products of the wave functions are permitted. These two
definitions are related with the shapes of ciliates and flagellates.
This might be a clue which could solve the problem of the
universality class of the swimming ways of micro-organisms.

In the next section, we review our previous papers \cite{1} about
ciliated and flagellated motions in the two dimensional fluid when
the Reynolds number $R \ll 1$. In Section 3, we consider the
area-preserving algebra which appears in the two dimensional fluid.
The appeared algebras are larger than the usual $w_{1+\infty}$ and
$W_{1+\infty}$ algebras. We give a free field representation of this
extended $W_{1+\infty}$ algebra. In Section 4, we show that there are
special state vectors in a representation of $W_{1+\infty}$ algebra.
These state vectors express the shape of micro-organisms. We also
consider about the hermitian conjugate and the inner product of these
state vectors and it is clarified there are only two ways to define
them. These two ways correspond to the shapes of ciliates and
flagellates. Last section is devoted to summary and discussion.

\section{The ciliated and flagellated motions in the two dimensional
fluid}

The micro-organisms with the length scale $L \ll 1$, the Reynolds
number $R$ satisfies $R \ll 1$, so that the hydrodynamics in this
case leads to the following equations of motion for the
incompressible fluid:
\begin{equation}
  \nabla \cdot  {\bf v} = 0\ ,    \label{2}
\end{equation}
and
\begin{equation}
  \Delta {\bf v}  = \frac{1}{\mu} \nabla p\ ,      \label{3.1}
\end{equation}
or equivalently
\begin{equation}
  \Delta (\nabla \times {\bf v}) = 0\ ,        \label{3.2}
\end{equation}
where $p$ is the pressure and ${\bf v}(x)$ is the velocity field of
the fluid. In two dimensions, the surface of a ciliate (flagellate)
becomes a closed (open) string and its position can be described by a
complex number
\begin{equation}
Z = x^1 + ix^2 = Z(t; \theta)\ ,
\end{equation}
with $-\pi \leq \theta \leq \pi$. In the sticky fluid of $\mu\neq 0$,
there is no slipping between the surface of a micro-organisim and the
fluid, namely, we have the matching condition
\begin{equation}
 {\bf v}({\bf x}={\bf X}(t; \xi))= \dot{{\bf X}}(t; \xi) \ .   \label{4.1}
\end{equation}

The ciliated motion can be viewed as a small but time-dependent
deformation of a unit circle in a properly chosen scale,
\begin{equation}
 Z(t, \theta) = s + \alpha(t, s)\ ,       \label{5}
\end{equation}
where $s=e^{i\theta}$ and $\alpha(t, s)$ is arbitrary temporally
periodic function with period $T$ satisfying $|\alpha(t, s)| \ll 1$
with $-\pi \leq \theta \leq \pi$. The complex representation of the
velocity vector $v_\mu$ can be denoted as
\begin{eqnarray}
&\ & v^z = 2v_{\bar{z}}(z, \bar{z})
= (v_1 +iv_2)(z, \bar{z})  \label{5.1}    \\
&\ & v^\zbar = 2v_z(z, \bar{z})
= (v_1 - iv_2)(z, \bar{z}).  \label{5.2}
\end{eqnarray}
By estimating the translational and rotational flows at spacial
infinity caused by the deformation of the cilia, we have obtained
$O(\alpha^2)$ expression of the net translationally swimming velocity
$v_{T}^{(\rm cilia)}$ of the ciliated micro-organism as follows:
\begin{eqnarray}
\lefteqn{ 2v_{T}^{({\rm cilia})}  =  -\dot{\alpha}_0 (t)} \nonumber   \\
      &  & + \sum_{n \leq 1} n (\dot{\alpha}_n \alpha_{-n+1}
      - \overline{\dot{\alpha}}_n \alpha_{n-1} - \overline{\dot{\alpha}}_n
      \overline{\alpha}_{-n+3})
- \sum_{n>1} n \dot{\alpha _n}
      \overline{\alpha}_{n-1}\ ,         \label{18}
\end{eqnarray}
where $\alpha_n (t)$ is defined by $\alpha (t, s) =
\sum_{n=-\infty}^{+\infty} \alpha_n (t) s^n$. On the other hand, the
net angular momentum $v_{R}^{({\rm cilia})}$ gained by the
micro-organism from the fluid becomes
\begin{eqnarray}
 \lefteqn{ 2v_{R}^{({\rm cilia})}  = -{\rm Im} \left\{ \dot{\alpha}_1 (t)
 \right.}   \nonumber      \\
    &  & - \left. \sum_{n \leq 1} n (\dot{\alpha}_n \alpha_{-n+2}
    - \overline{\dot{\alpha}}_n \alpha_{n} - \overline{\dot{\alpha}}_n
    \overline{\alpha}_{-n+2})
+  \sum_{n>1} n \dot{\alpha _n}
    \overline{\alpha}_n \right\}.                  \label{19}
\end{eqnarray}
The net translation and rotation resulted after the period $T$ come
from $O(\alpha^2)$ terms since the $O(\alpha)$ terms cancel after the
time integration over the period.

Micro-organisms swimming using a single flagellum can be viewed as an
open string with two endpoints, H and T, where H and T represent the
head and the tail-end of a flagellum, respectively. Our discussion
will be given by assuming that the distance between H and T is
time-independent and is chosen to be 4 in a proper length scale. This
assumption can be shown to be valid for the flagellated motion by
small deformations in the incompressible fluid. Then, at any time
$t$, we can take a complex plane of $z$, where H and T are fixed on
$z=2$ and $-2$, respectively. This coordinate system $z$ can be
viewed as that of the space of {\it standard shapes} named by Shapere
and Wilczek \cite{3}. Time dependent, but small deformation of the
flagellate can be parametrized as
\begin{equation}
   Z(t, \theta) = 2(\cos \theta + i\sin \theta \alpha (t, \theta)) ,
   \label{20}
\end{equation}
where the small deformation $\alpha (t, \theta)$ can be taken to be a
real number\footnote{When $\alpha$ is taken to be a complex number,
the length of the flagellum is locally changeable at $O(\alpha)$. For
such an elastic flagellum, we have similar results to that of the
ciliated motion. In case of real $\alpha$, its length is locally
preserved at $O(\alpha)$, giving a non-elastic flagellum, which is
the more realistic one.} satisfying
\begin{equation}
  \alpha (t, \theta) = -\alpha (t, -\theta).        \label{21}
\end{equation}
Here, we parametrize the position of the flagellum twice, starting
from the endpoint T at $\theta = -\pi$, coming to the head H at
$\theta = 0$, and returning to T again at $\theta = \pi$. Motion of
the two branches corresponding to $-\pi \leq \theta \leq 0$ and
$\pi \geq \theta \geq 0$ should move coincidentally, which requires
the condition (\ref{21}). The Joukowski transformation
$z = z(w)= w + w^{-1}$, separates the two coincident branches in the
$z$ plane to form lower and upper parts of a unit circle in the $w$
plane, outside domain of which we are able to study the swimming
problem of the flagellate in a quite similar fashion to that of the
ciliate. The parametrization of our micro-organism in the $w$ plane
corresponding to Eq.(\ref{20}) is now
\begin{equation}
  W(t, \theta) = e^{i \theta}(1 + \alpha (t, \theta)) + O(\alpha^2).
       \label{22}
\end{equation}
Using the mode expansion satisfying Eq.(\ref{21}),
\begin{equation}
  \alpha (t, \theta) = \sum_{n=1}^{\infty} \alpha_n (t) \sin n\theta,
  \label{31}
\end{equation}
we are able to determine the net swimming velocity
${v_T}^{\rm(flagella)}$ gained by the flagellate motion of
micro-organism:
\begin{equation}
  2v_T ^{(\rm flagella)} = -i \dot{\alpha}_1
  - \sum_{m \geq 1} m \alpha_m \dot{\alpha}_{m+1} +
  \sum_{m \geq 2} m \alpha_m \dot{\alpha}_{m-1} ,   \label{34}
\end{equation}
On the other hand, the angular momentum ${v_R}^{(\rm flagella)}$ is
given by
\begin{equation}
   2{v_R}^{(\rm flagella)}  = - \frac{1}{2} \dot{\alpha}_2.   \label{35}
\end{equation}
After the time integration over the period $T$,
${v_R}^{(\rm flagella)}$ vanishes since in our first order
approximation, the length of the flagellum is fixed in the
incompressible fluid. Therefore the second order approximation
is necessary for the non-vanishing ${v_R}^{(\rm flagella)}$.

\section{The $w_{1+\infty}$ and $W_{1+\infty}$ algebras in
the two dimensional fluid}

The velocity field of fluid $\bfv (x)$ transforms the volume element
of fluid on $\bfx$ into $\bfx +\delta t \bfv$ in an infinitesimally
small time interval $\delta t$. If the fluid is imcompressible
(\ref{2}), the velocity field ${\bf v}$ generates area-preserving
diffeomorphism in two dimensions. Therefore the velocity field
${\bf v}$ is given by a scalar function $U(z,\zbar)$, which is called
a stream function,
\begin{equation}
\label{stream}
v_z=-\half\parz U(z,\zbar)  \ , \ \ \
v_\zbar=\half\parzbar U(z,\zbar)   \ .
\end{equation}
Since $\overline{v_z}=v_\zbar$, $U$ should be pure imaginary. The generators
of the area-preserving diffeomorphism are given by,
\begin{equation}
L_U=\parzbar U\parz - \parz U\parzbar\ ,
\end{equation}
and the commutation relations between the generators are expressed as
\begin{equation}
[L_U, L_V]=L_{[U,V]_{{\rm P.B.}}}\ .
\end{equation}
Here $[U,V]_{{\rm P.B.}}$ is the Poisson bracket defined by,
\begin{equation}
[U,V]_{{\rm P.B.}}=\parzbar U\parz V - \parz U\parzbar V\ .
\end{equation}
If we choose the basis of $U$ by $\{z^n\zbar^m\}$ ($n$ and $m$ are
integers) and define $L_{nm}$ by $L_{nm}\equiv L_{U=z^n\zbar^m}$, we
obtain $w_{1+\infty}$ algebra:
\begin{equation}
\label{winf}
[L_{nm}, L_{kl}]=-(nl-mk)L_{n+k-1\ m+l-1}\ .
\end{equation}
If we solve the equation of motions (\ref{2}) and (\ref{3.1}) or
(\ref{3.2}) when the Reynolds number $R\ll 1$, the stream function
$U$ contains $z\bar{z}^k$, $z^k \bar{z}$, $\ln z$, $\ln \bar{z}$,
$z{\ln}z\bar{z}$ and $\bar{z}{\ln}z\bar{z}$ terms ($k$ : integer).
Since $U$ contains $\ln z$ and $\ln \bar{z}$ terms, an algebra larger
than the usual $w_{1+\infty}$ algebra in Eq.(\ref{winf}) is
generated. The algebra so obtained consists of
$L_{(l, m, n)} \equiv L_{z^l \bar{z}^m (\ln z \bar{z})^n}$, and
$M \equiv L_{\ln z - \ln \bar{z}}$; They satisfy
\begin{eqnarray}
\label{xw1}
[L_{(l, m, n)}, \ L_{(p, q, r)}] & = & -(lq-mp)\,
L_{(l+p-1,\,  m+q-1,\,  n+r)} \nonumber  \\
    & & +(mr+np-lr-nq)\, L_{(l+p-1, \, m+q-1, \, n+r-1)} \nonumber  \\
  & & + c \, (m-l)\,
  \delta _{l+p, \, 0} \, \delta _{m+q, \, 0} \,
\delta _{n+r, \, 0}    \ \  , \\
  \left[ M, \ M\right] & = & 0 \ \ ,
\end{eqnarray}
and
\begin{eqnarray}
\label{xw2}
[M, \ L_{(l, m, n)}] & = & -(l+m)\, L_{(l-1, m-1, n)}
- 2n \, L_{(l-1, m-1, n-1)}  \nonumber \\
& & \ \ \ \ + \frac{1}{2} c \, \delta_{l, \, 0} \, \delta_{m, \, 0} \,
\delta_{n, \, 0}.
\end{eqnarray}
In the above expression we add the central charge $c$, corresponding
to the possible central extension of the algebra in which the Jacobi
identities are kept to hold and the generators are understood to be
properly redefined. We call the algebra in Eqs.(\ref{xw1}) and
(\ref{xw2}) as ``extended'' $w_{1+\infty}$ algebra.

The reason why the $\ln z$ or $\ln \bar{z}$ is permitted in the
stream function $U  (z, \bar{z})$ is that the existing singularities
at $z=0$ can be hidden inside the body of the micro-organism itself.
Therefore, if we are not interested in the circulation flow of the
fluid (topological flow) around the micro-organism, we can ignore the
logarithmic contribution in $U$. In that case, the algebra becomes
$w_{1+ \infty}$ in Eq.(\ref{winf}).

$W_{1+\infty}$ algebra is obtained by \lq\lq quantizing''
$w_{1+\infty}$ algebra, {\it i.e.}, replacing $\zbar$ and the Poisson
bracket in $w_{1+\infty}$ algebra with $\hbar\parz$ and the
commutator, respectively. Then the generators $\hat L_{nm}$
corresponding $L_{nm}$ in $w_{1+\infty}$ algebra are given by
\begin{equation}
\hat L_{nm}\equiv \hbar^{m-1}z^{n-m}D^m, \ \ \ D\equiv z\parz\ ,
\end{equation}
and their commutation relations have the following forms:
\begin{eqnarray}
[\hat L_{nm}, \hat L_{kl}] &=& \sum_{j=1}^\infty \left\{
{(m+1)(k-l)^j \over B(j+1,m-j+1)}
-{(l+1)(n-m)^j \over B(j+1,l-j+1)} \right\} \nonumber \\
&\ &\hskip 2cm
\times\hbar^{j-1}\hat L_{n+k-j\ m+l-j} \label{wone} \\
&=& -(nl-km)\hat L_{n+k-1\ m+l-1}+O  (\hbar)\ . \label{wone2}
\end{eqnarray}
Here $B(p,q)$ is the beta function:
$B(p,q)=\Gamma (p)\Gamma (q)/\Gamma (p+q)$. The sum in
Eq.(\ref{wone}) is finite if $m\geq 0$ and $l\geq 0$ since
$1/B(p,q)=0$ when $p$ or $q$ is a negative integer while $p+q$ being
a positive integer. Usually $m$ in $\hat L_{nm}$ is non-negative
integer but we need to include negative one here since there often
appears negative power of $\zbar$ in the stream function and $\zbar$
corresponds to $\hbar \parz$. It is straightforward to extend the
usual $W_{1+\infty}$ algebra to the algebra containing the operators
with negative power of $D$ if we understand that the formula:
$D^mz^n=z^n(D+n)^m$ holds also for negative $m$. Eq.(\ref{wone2})
tells that $w_{1+\infty}$ algebra is reproduced in the limit of
$\hbar\rightarrow 0$.

It is straightforward to quantize the extended $w_{1+\infty}$ algebra
in Eqs.(\ref{xw1}) and (\ref{xw2}) by defining,
\begin{eqnarray}
\label{xW}
\hat L_{(l, m, n)}&\equiv& \hbar^{m-1}z^{l-m}D^m
(\ln D)^n\ ,\nonumber \\
\hat M&\equiv& 2\ln z - \ln D\ .
\end{eqnarray}
We call the algebra generated by $\hat L_{(l, m, n)}$ and $\hat M$ as
``extended'' $W_{1+\infty}$ algebra. We can also construct the
central extension of this algebra by using free fields $b(z),\ c(z)$:
\begin{eqnarray}
b(z)&=&\sum_{l:{\rm integer}}b_lz^{-l+s}\ , \ \ \
c(z)=\sum_{l:{\rm integer}}c_lz^{-l-s-1} \ , \nonumber \\
&\ &[b_l,\ c_k]_\varepsilon =\delta_{l+k,0}\ .
\end{eqnarray}
Here $\varepsilon=+$ $(-)$ if $b(z),\ c(z)$ are fermions (bosons) and
$[\ ,\ ]_{+\,(-)}$ is (anti-)commutator. The generators
${\cal L}_{(l, m, n)}$, which corresponds to $\hat L_{(l, m, n)}$, in
the central extended algebra are given by,
\begin{eqnarray}
{\cal L}_{(l, m, n)}&=&\varepsilon\oint{dz \over 2\pi i}
:c(z)\hat L_{(l, m, n)}b(z): \nonumber \\
&=&\varepsilon\sum_{k:{\rm integer}}\hbar^{m-1} :c_{k+l-m}
(k+s)^m\{\ln (k+s)\}^n b_{-k}:\ .
\end{eqnarray}
Here $:\ :$ means normal-ordering. In order that $\{\ln (k+s)\}^n$
and $(k+s)^m$ for negative $m$ are well-defined, $s$ should not be an
integer. There can be two choices to define $\ln (k+s)$ when $k+s<0$:
$\ln (k+s)\equiv \ln |k+s| +i\pi$ or
$\ln (k+s)\equiv \ln |k+s| -i\pi$. The commutator is given by
\begin{eqnarray}
[{\cal L}_{(l, m, n)}, \ {\cal L}_{(p, q, r)}]
&=&-(lq-mp){\cal L}_{(l+p-1, m+q-1, n+r)} \nonumber \\
&\ & +(mr+np-lr-nq){\cal L}_{(l+p-1, m+q-1, n+r-1)}
\nonumber \\
&\ & +\Bigl( O(\hbar)\ {\rm operator\ terms} \Bigr)
\nonumber \\
&\ & +\varepsilon \delta_{l-m+p-q,0} \hbar^{m+q-2}
\Bigl( \sum_{k>0}-\sum_{k>l-m} \Bigr) \nonumber \\
&\ & \hskip 1cm \times (k+s)^q(k+s-l+m)^m \nonumber \\
&\ & \hskip 1cm \times \{\ln (k+s)\}^r
\{\ln (k+s-l+m)\}^n \ .
\end{eqnarray}
It is not so straightforward to obtain an operator ${\cal M}$
corresponding to $\hat M$ in Eq.(\ref{xW}) since $\hat M$ contains
$\ln z$ term. An expression for ${\cal M}$ can be obtained by
bosonizing free fields $b(z),\ c(z)$ \cite{10}:
\begin{eqnarray}
b(z)&=&:\e^{\phi(z)}:\ , \ \ \ \ c(z)=:\e^{-\phi(z)}:\ \ \ \ \
(b,\ c\ :{\rm fermions}) \\
b(z)&=&:\e^{\phi(z)}:\eta(z)\ , \
\ c(z)=:\e^{-\phi(z)}:\partial\zeta(z) \ \
(b,\ c\ :{\rm bosons})
\end{eqnarray}
\begin{equation}
\phi(z)=q+\alpha_0\ln z + ({\rm oscillating\ terms}) \ , \ \ \ \
[\alpha_0,\ q]=1\ .
\end{equation}
Then we find
\begin{equation}
[q, {\cal L}_{l,m,n}]=m{\cal L}_{l-1,m-1,n}
+n{\cal L}_{l-1,m-1,n-1}\ .
\end{equation}
Therefore if we define ${\cal M}$ by
\begin{eqnarray}
{\cal M}\equiv -2q + {\cal L}_{0,0,1}\ ,
\end{eqnarray}
we obtain
\begin{eqnarray}
[{\cal M}, \ {\cal L}_{l,m,n}]&=&
-(l+m){\cal L}_{l-1,m-1,n}-2n{\cal L}_{l-1,m-1,n-1} \nonumber \\
&\ & +\Bigl( O(\hbar)\ {\rm and\ {\it c}\
number\ terms}\Bigr)\ .
\end{eqnarray}

\section{Wave Functions for the Shape of Micro-Organisms}

In the following, we consider the $W_{1+\infty}$ algebra whose
central charge vanishes. It is because we are mainly interested in
the classical motion of micro-organisms and we cannot take the
classical limit: $\hbar\rightarrow 0$ if the central charge does not
vanish. Since the operators of $W_{1+\infty}$ is given by $z$ and
$\parz$ when the central charge vanishes, a set of the representation
is given by functions of $z$ and basis vectors are given by
$\{z^n,\ n:{\rm integer}\}$. This representation is {\it not} the
usual highest weight representation. Especially, if we regard
$\hat L_{11}$ with the Hamiltonian of the system, the energy becomes
unbounded below. When we consider the motion of fluid, however,
$\hat L_{11}$ is not the Hamiltonian and there is not any problem.
This representation is very useful for the intuitive understanding of
the swimming motion of micro-organisms as we will see in the
following.

We now consider the commutator between $\hat L_{nm}$ and
$z=\hbar \hat L_{10}$:
\begin{equation}
[\hat L_{nm}, z]=\hbar^{m-1}z^{n-m+1}\{(D+1)^m-D^m\}\ .
\end{equation}
By assuming that a state vector of the representation is given by
$\e^{{f(z) \over \hbar}}$, we obtain
\begin{equation}
\label{delz}
[\hat L_{nm}, z]\e^{{f(z) \over \hbar}}
=\Bigl\{ mz^n (\parz f(z))^{m-1}+O (\hbar)\Bigr\}\e^{{f(z) \over \hbar}}.
\end{equation}
On the other hand, the commutator of $L_{nm}$ in the classical
$w_{1+\infty}$ algebra and $z$ is given by,
\begin{equation}
\label{delz2}
[L_{nm}, z]=mz^n\zbar^{m-1}\ .
\end{equation}
By comparing (\ref{delz}) with (\ref{delz2}), we find that
$[\hat L_{nm}, z]$ coincides with $[L_{nm}, z]$ in the classical
limit $\hbar \rightarrow 0$ if the following equation holds
\begin{equation}
\label{shape}
\zbar=\parz f(z)\ .
\end{equation}
If we choose a special class of functions $f(z)$, the points
satisfying Eq.(\ref{shape}) can form a closed line in $z$-plane or
Riemann sphere. The shapes of (the surface of) micro-organisms are
always given by the equation in the form of Eq.(\ref{shape}). For
example, since the basic shape of a ciliate with unit radius in
Eq.(\ref{5}) is given by $\zbar = z^{-1}$, $f(z)$ has the following
form:
\begin{equation}
\label{cilia}
f_{{\rm cilia}}(z)=\ln z\ .
\end{equation}
On the other hand, the basic shape of a flagellate in Eq.(\ref{20})
is given by a real axis $\zbar = z$ and we find the corresponding
$f(z)$;
\begin{equation}
\label{flagellum}
f_{{\rm flagellum}}(z)=\half z^2\ .
\end{equation}
If Eq.(\ref{shape}) gives the shape of a micro-organism,
$[\hat L_{nm}, z]$ coincides with $[L_{nm}, z]$ in the classical limit
$\hbar \rightarrow 0$ just on the surface of micro-organisms.
In this sense, we can regard that the state vector
$\e^{{f(z) \over \hbar}}$ represents the shape of a micro-organism
and we call the state vector $\e^{{f(z) \over \hbar}}$ and
Eq.(\ref{shape}) as the ``wave function'' and ``shape'' equation of
micro-organism, respectively.

In order to verify that the shape equation (\ref{shape}) really
expresses the shape of micro-organism, we compare the operations of
$\hat L_{nm}$ and $L_{nm}$ on the shape equation (\ref{shape}). If we
operate $1+\epsilon\hat L_{nm}$ on a wave function
$\e^{{f(z) \over \hbar}}$, we obtain,
\begin{equation}
\label{defo}
(1+\epsilon\hat L_{nm})\e^{{f(z) \over \hbar}}
=\Bigl\{1+\epsilon\hbar^{-1}\Bigl(z^n(\parz f(z))^m+O (\hbar)
\Bigr) \Bigr\}\e^{{f(z) \over \hbar}}.
\end{equation}
Eq.(\ref{defo}) tells that $f(z)$ is changed by
\begin{equation}
f(z)\rightarrow f(z)+\epsilon z^n(\parz f(z))^m +O (\epsilon^2)
+O(\hbar)\ .
\end{equation}
Therefore the shape equation (\ref{shape}) is also changed by
\begin{eqnarray}
\label{sdef}
\zbar &=& \parz \Bigl\{f(z)+\epsilon z^n(\parz f(z))^m+O (\epsilon^2)
+ O(\hbar) \Bigr\} \nonumber \\
&=& \parz f(z)+m\epsilon z^n(\parz f(z))^{m-1}\parz^2 f(z)
+n\epsilon z^{n-1}(\parz f(z))^m +O (\epsilon^2) +O(\hbar)
\nonumber \\
&=& \parz f(z)+m\epsilon z^n\zbar^{m-1}\parz^2 f(z)
+n\epsilon z^{n-1}\zbar^m +O (\epsilon^2) + O(\hbar)\ .
\end{eqnarray}
On the other hand, the classical operator $\epsilon L_{nm}$
transforms $z$ and $\zbar$ as
\begin{equation}
z\rightarrow z+\epsilon mz^n\zbar^{m-1}\ , \ \ \
\zbar \rightarrow \zbar-\epsilon nz^{n-1}\zbar^m\ .
\end{equation}
Therefore the shape equation (\ref{shape}) should be changed by the
operation of $L_{nm}$,
\begin{eqnarray}
\label{sdef2}
\zbar-\epsilon nz^{n-1}\zbar^m&=&\parz f(z+\epsilon mz^n\zbar^{m-1})
\nonumber \\
&=& \parz f(z)+\epsilon mz^n\zbar^{m-1}\parz^2 f(z) + O  (\epsilon^2) \ .
\end{eqnarray}
The above equation (\ref{sdef2}) is identical with Eq.(\ref{sdef}).
This tells that $\hat L_{nm}$ exactly reproduces the deformation of
the shape of micro-organisms in the classical limit:
$\hbar\rightarrow 0$.

In the following, we define the hermitian conjugate of wave function
$\Phi(z)$. The complex conjugate $\bar \Phi(\bar z)$ of $\Phi(z)$ is,
of course, a function of $\zbar$. We define the hermitian conjugate,
which is a function of $z$ by
\begin{equation}
\Phi^\dagger (z)=\bar \Phi (\parz f_0(z))\ .
\end{equation}
Here $f_0(z)$ is a fixed function which specifies a proper shape by
the shape equation (\ref{shape}): $\zbar =\parz f_0(z)$. Since
$(\Phi^\dagger (z))^\dagger$ should be $\Phi (z)$, $f_0(z)$ should
satisfies the equation
$z=\parzbar \bar f_0(\zbar)|_{\zbar =\parz f_0(z)}$ for arbitrary
$z$. There are only two classes of solutions. The solutions are given
by $f_{{\rm cilia}}(z)$ in
Eq.(\ref{cilia}) and $f_{{\rm flagellum}}(z)$ in Eq.(\ref{flagellum})
up to rotation, finite translation and finite scale transformation.
It might be remarkable that there are only two classes in $f_0(z)$.
This might be one of the reasons why there are just two classes of
the swimming ways, {\it i.e.}, ciliated and flagellated motions in
two dimensions.

We now define the inner product $<\Psi|\Phi>$ of two wave functions
$\Psi(z)$ and $\Phi(z)$ by
\begin{equation}
\label{inner}
<\Psi|\Phi>\equiv \int_C ds \Psi^\dagger(z(s)) \Phi(z(s))\ .
\end{equation}
Here $s$ parametrizes $C$ which is a contour deformable from the closed line
$\zbar = \parz f_0(z)$ without crossing any singularity in
$\Psi^\dagger(z) \Phi(z)$. Then the inner product $<\Phi|\Phi>$ is
positive-definite since
\begin{eqnarray}
<\Phi|\Phi>&=&\oint_C ds \Phi^\dagger(z(s)) \Phi(z(s)) \nonumber \\
&=& \oint_{\zbar = \parz f_0(z) \ {\rm line}} ds
\Phi^\dagger(z(s)) \Phi(z(s)) \nonumber \\
&=& \oint_{\zbar = \parz f_0(z) \ {\rm line}} ds
\bar\Phi(\bar z(s)) \Phi(z(s)) \nonumber \\
&>& 0\ .
\end{eqnarray}
By using the definition of the inner product (\ref{inner}), we can
define the hermitian conjugate of the operators $\hat L_{nm}$ so as
to
\begin{equation}
<\hat L_{nm} \Psi|\Phi> = <\Psi| \hat L_{nm}^\dagger \Phi>\ .
\end{equation}
If we choose $f_0(z)=f_{{\rm cilia}}(z)=\ln z$, the hermitian
conjugate is given by
\begin{equation}
\hat L_{nm}^\dagger = \sum_{j=0}^\infty
{\hbar^j (m+1) (m-n)^j \over B(m-j+1, j+1)} \hat L_{2m-n+j\ m-j}\ .
\end{equation}
On the other hand, if we choose
$f_0(z)=f_{{\rm flagellum}}=\half z^2$, we obtain
\begin{equation}
\hat L_{nm}^\dagger = \sum_{j=0}^\infty
{\hbar^j (-1)^{m+j} (m+1) (m-n-1)^j \over B(m-j+1, j+1)}
\hat L_{n+j\ m-j}\ .
\end{equation}

By using the above formulation, we consider the swimming motion of
the micro-organisms in the following.

If the velocity field $v^z(t)=2v_\zbar(t)$ is given by
\begin{eqnarray}
v^z(t)&=&\sum_{n,\ m} m\alpha_{nm}(t)z^n\zbar^{m-1}
\nonumber \\
v^\zbar(t)&=&\overline{v^z(t)} \nonumber \\
&=&\sum_{n,\ m} m\bar\alpha_{nm}(t)\zbar^n z^{m-1}
\end{eqnarray}
the corresponding operator $\hat L(t)$ in $W_{1+\infty}$ algebra has
the following form:
\begin{equation}
\label{Lt}
\hat L(t)=\sum_{n,\ m\neq 0} {1 \over 2}
(\alpha_{nm}(t)-\bar\alpha_{mn}(t))\hat L_{nm}\ .
\end{equation}
Note that $\bar\alpha_{mn}=-\alpha_{nm}$ since the stream function
$U$ in Eq.(\ref{stream}) is pure imaginary. $\hat L(t)$ is a quantum
version of $U$. $\hat L(t)$ generates the time-development of a wave
function $\Phi (z)$ by
\begin{equation}
\label{tdev}
\Phi (z,t)=T\e^{\int_0^tdt \hat L(t)}\Phi (z)\ .
\end{equation}
Here $T$ means the time-ordering. It is important to recognize the
following: The ``Hamiltonian'' $i\hat L(t)$ can be obtained from the
stream function $iU(z,\zbar)$ by replacing $\zbar$ with $\hbar\parz$.
Therefore what we are doing here is the quantum mechanics
corresponding to the classical mechanics whose Hamilton equation is
${\partial z \over \partial t}=\parzbar U(z,\zbar)$ in
Eq.(\ref{stream}). The expectation value of any operator $\co$
at time $t$ is given by, ($AT$ means the anti-time ordering.)
\begin{equation}
<\co (t)>={<\Phi|AT\e^{-\int_0^tdt \hat L(t)} \co
T\e^{\int_0^tdt \hat L(t)} |\Phi> \over <\Phi|\Phi>} \ .
\end{equation}
In other words, the time development of operator $\co$ is given by
$\co(t)=AT\e^{-\int_0^tdt \hat L(t)} \co T\e^{\int_0^tdt \hat L(t)}$.
Especially velocity operator $\hat v^z(t)$ is given by
\begin{eqnarray}
\label{vope}
\hat v^z(t)&=&{d \over dt}\Bigl\{
AT\e^{-\int_0^tdt \hat L(t)} z
T\e^{\int_0^tds \hat L(s)}\Bigr\} \nonumber \\
&=&-[\hat L(t),z]+[\int_0^t ds \hat L(s), [\hat L(t),z]]
+O ((\hat L(t))^3)\ .
\end{eqnarray}
In case of ciliated motion in the fluid with the Reynolds number
$R\ll 0$, the velocity field $v^z$ is given by \cite{1},
\begin{equation}
v^z=\sum_{n\geq -1}\dot \alpha_{-n}z^{-n}+
\sum_{n\geq 1}\{\dot\alpha_{n+1}
-(n-1)\overline{\dot \alpha}_{-n+1}\}\zbar^{-n-1}
+\sum_{n\geq 0}n\overline{\dot \alpha}_{-n}\zbar^{-n-1}z
\end{equation}
when the ciliated motion is given by a small but time-dependent
deformation of a unit circle in Eq.(\ref{5}). By using Eq.(\ref{Lt}),
we find that  $\hat L(t)$ is given by
\begin{eqnarray}
\hat L(t) &=& \sum_{n\geq -1}\dot \alpha_{-n}\hat L_{-n\ 1}-
\sum_{n\geq 1}{1 \over n}\{\dot \alpha_{n+1}
-(n-1)\overline{\dot \alpha}_{-n+1}\}\hat L_{0\ -n} \nonumber \\
 &\ & -\sum_{n\geq -1}\overline{\dot \alpha}_{-n}\hat L_{1\ -n}+
\sum_{n\geq 1}{1 \over n}\{\overline{\dot \alpha}_{n+1}
-(n-1)\dot \alpha_{-n+1}\}\hat L_{-n\ 0}\ .
\end{eqnarray}
Then by operating $\hat v^z(t)$ in Eq.(\ref{vope}) on the wave
function $\e^{f_{{\rm cilia}} \over \hbar}=z^{{1 \over \hbar}}$ which
expresses the basic shape of ciliates, {\it i.e.}, a unit circle, and
by extracting constant part $v_{T}^{(\rm cilia)}$, which gives a net
translationally swimming velocity of the micro-organism, we have
succeeded in obtaining the result of Eq.(\ref{18}) from the viewpoint
of $W_{1+\infty}$ algebra.

\section{Summary}

In this paper, the swimming problem of the micro-organisms was analyzed
in two dimensions from the algebraic viewpoint by using
$w_{1+\infty}$ and/or $W_{1+\infty}$ algebra. Semi-classical
equivalence between the representations of the algebra of
area-preserving diffeomorphisms $w_{1+\infty}$ and its ``quantized''
version $W_{1+\infty}$ leads to the wave functions which express the
shape of micro-organisms. In order to define consistently the
hermitian conjugate and the inner products of the wave functions, we
have found that there exist only two different ways corresponding to
the shapes of ciliates and flagellates. By using the ``quantized''
algebra $W_{1+\infty}$ and their wave functions, the swimming
velocity of the ciliates can be reproduced. The formulation proposed
in this paper using $W_{1+\infty}$ algebra and the wave functions is
the quantum mechanics of the fluid dynamics where the stream function
plays the role of the Hamiltonian. In the swimming problem of
micro-organisms in two dimensional fluid, the area-preserving
algebras can be larger than the usual $w_{1+\infty}$ and
$W_{1+\infty}$ algebras. We have studied these extended
$w_{1+\infty}$ and $W_{1+\infty}$ algebras and given a free field
representation of the latter.

\section*{Acknowledgements}
This work is partly supported by Grant-in-Aid for Scientific Research
from the Ministry of Education, Science and Culture (No.06221229).

\end{document}